\documentclass[12pt]{iopart}

\usepackage{graphicx}
\usepackage{color}

\begin{document}

\title{Quantum machine learning and quantum biomimetics: A perspective}

\author{Lucas Lamata}

\address{Departamento de F\'isica At\'omica, Molecular y Nuclear, Universidad de Sevilla, 41080 Sevilla, Spain}
\vspace{10pt}
\begin{indented}
\item[]April 2020
\end{indented}

\begin{abstract}
Quantum machine learning has emerged as an exciting and promising paradigm inside quantum technologies. It may permit, on the one hand, to carry out more efficient machine learning calculations by means of quantum devices, while, on the other hand, to employ machine learning techniques to better control quantum systems. Inside quantum machine learning, quantum reinforcement learning aims at developing ``intelligent'' quantum agents that may interact with the outer world and adapt to it, with the strategy of achieving some final goal. Another paradigm inside quantum machine learning is that of quantum autoencoders, which may allow one for employing fewer resources in a quantum device via a training process. Moreover, the field of quantum biomimetics aims at establishing analogies between biological and quantum systems, to look for previously inadvertent connections that may enable useful applications. Two recent examples are the concepts of quantum artificial life, as well as of quantum memristors. In this Perspective, we give an overview of these topics, describing the related research carried out by the scientific community.
\end{abstract}

\vspace{2pc}

\section{Introduction}

The field of quantum technologies has experienced a significant boost in the past five years. The interest of multinational companies such as Google, IBM, Microsoft, Intel, and Alibaba in this area has increased the worldwide competition and available funding not only from these firms, but also from national and supranational governments, such as the European Union~\cite{Google1,EU}. Most of these companies have also a strong focus on machine learning, which is a wider-deployed technology, as it is currently more advanced. Increases in the past decade in computing power, via, e.g., Graphical Processing Units (GPUs), and the extensive availability of data in internet, have enabled a significantly good performance of deep learning. Among others, image recognition, language translation, search engines, automated cars, and personal assistants, have become widespread nowadays~\cite{HintonNature,RussellAI}. 

Given that quantum systems are well described by linear algebra, and this also applies to several machine learning algorithms, it was only a matter of time that these two fields merged in a new area, namely, quantum machine learning. An exponentially-growing amount of papers has appeared in this area in the past three to four years~\cite{WOK}, and already several books and reviews on this field have been published~\cite{PeterWittek,Biamonte2017Review,DunjkoReview,SchuldQML,SchuldBook,SchuldQNN,DunjkoWittek}. Existing algorithms include quantum solvers of linear systems of equations~\cite{HHL}, quantum support vector machines~\cite{QSVM}, quantum principal component analysis~\cite{QPCA}, quantum gradient estimation~\cite{QGE}, and distributed secure quantum machine learning~\cite{ShengDSQML}, among others. Further research areas inside quantum machine learning include quantum reinforcement learning~\cite{DongQRL,PaparoPRX,DunjkoPRL,LamataQRL,CardenasQRL,AlbarranQRL,ShangYuQRL,AlbarranMLST}, quantum autoencoders~\cite{AlanAutoencoder,KimAutoencoder,LamataAutoencoder,TischlerAutoencoder,DingAutoencoder,AminAutoencoder}, quantum neural networks~\cite{SchuldQNN,KimAutoencoder,TacchinoQNN,TorronteguiQNN,AlanQNN,MikelQNN1,MikelQNN2,UnaiSupervisedLearning}, as well as quantum annealing~\cite{DWaveNature,VenturelliQA,RieffelQA}, quantum Boltzmann~\cite{AminMelkoQBM,WiebeQBM,PerdomoPRX} and quantum Helmholtz machines~\cite{PerdomoHelmholtz,PerdomoHelmholtz2}. Another research avenue complementary to the former ones, consists in employing machine learning algorithms to better control quantum systems, reduce gate errors, increase state preparation fidelities, and compute quantum phases of matter~\cite{Ghosh1,Ghosh2,SanzGenetic,RuiGenetic,Carrasquilla,MehtaPR,LiuVD,CiracRMP}. Moreover, inside this emerging area, a growing amount of papers in the field of machine learning for better understanding quantum systems can be also classified in reinforcement learning~\cite{MarquardtRL,BukovPRX,Bukov2,MelnikovQL,MelnikovBell,Jelena,PaternostroThermo,MelnikovComm,Lorch,XinWangnpjQuantum,XinWang2,EisertML,LiangJiang}, supervised learning~\cite{PaternostroSupervised,Tomamichel,Milburn,Filippov,MelnikovQUTE,MelnikovNJP,JacobTaylorNPJ,EndresPRL,CarleoNatPhys}, and unsupervised learning~\cite{Cappellaro,Sciarrino} works. For a recent review on this specific field, see~\cite{CarrasquillaReview}. Finally, another interesting topic that has appeared recently is the one of quantum-inspired algorithms, namely, classical machine learning algorithms that get inspiration from quantum algorithms~\cite{Tang,Arrazola}.

Another field that aims at connecting two previously disconnected research areas such as quantum technologies and biological systems~\cite{al1,alReview,al2} is the one of quantum biomimetics. This has some overlap with quantum machine learning in topics such as, e.g., quantum neural networks, which are also biomimetic: They consist of a quantization of neural networks that, in turn, get inspiration from the human brain, a biological system. Quantum artificial life is an area inside quantum biomimetics in which the aim is to design quantum individuals that can self-replicate in a quantum fashion, namely, without breaking the no-cloning theorem~\cite{qubiom1,qubiom2,qubiom3,WootersZurek}. They can also mutate, and evolve with a quantum Darwinian process. Even though this is still a field in its infancy, we expect that it could have connections with quantum game theory~\cite{EisertQG} and could allow one for encoding optimization problems, being this way connected again with the quantum machine learning field. Related research has been carried out by diverse groups~\cite{qe1,qe2,qe3,qe4}. A second area inside quantum biomimetics is the one of quantum memristors~\cite{qmem1,qmem2,qmem3,qmem4}. These are devices with processing power, memory, and quantum character, inspired in the classical memristor, the fourth element in electronic engineering in addition to the resistor, capacitor, and inductor~\cite{DiVentraMemristors}. A classical memristor has processing capacity at the same time as memory, and has been conjectured to be able to solve NP problems in polynomial time. Classical memristors may provide a novel paradigm for computing, with a neuromorphic architecture, and, equivalently, quantum memristors represent the basic building block for a neuromorphic quantum computer. They have also a close connection to quantum neural networks, and may provide a platform to quantum simulate non-Markovian quantum open systems.

In this Perspective, we will revise the fields of quantum machine learning and quantum biomimetics, without the goal of being exhaustive, and focusing mainly on our own research of the past six years, as well as related research by other groups. Our motivation is to give our vision of what has been the development of these two fields in the past few years, and what could be their future evolution. Even though these two areas may seem rather disconnected at first sight, they have a strong relationship, as described for example in Ref.~\cite{alReview} for the classical cases: living systems, either natural or artificial, have often the ability to learn. Being this article also a Perspective, it is naturally more biased towards my own research, and that of my collaborators at the QUTIS Group in University of the Basque Country~\cite{QUTIS}, which has considered these two fields in parallel and with multiple links. As examples of this, one can mention the international conference we coorganized in Bilbao in March 2018, on ``Quantum Machine Learning \& Biomimetic Quantum Technologies'', as well as the Special Issue we edited on ``Quantum Machine Learning and Bioinspired Quantum Technologies''~\cite{SI_QML}. We will first start with Section~\ref{SectionQML}, on quantum machine learning, in which we will describe two emerging topics inside this area, namely, quantum reinforcement learning, and quantum autoencoders. In Section~\ref{QAL}, on quantum biomimetics, we will revise other two topics, i.e., quantum artificial life, both as a proposal and its experimental realization in the IBM cloud quantum computer, and quantum memristors. We will then give our conclusions in Section~\ref{Conclusions}.

\section{Quantum machine learning\label{SectionQML}}

Machine learning is one of the fields with a larger impact in computer science nowadays~\cite{RussellAI}. Its influence is widespread along a large number of technologies and economies, and this will only grow in the future. One of the reasons for this success is the so-called deep learning technology~\cite{HintonNature}. This kind of approach, based on neural networks with several layers, and the backpropagation algorithm, has existed since the 80's, but only recently has started to work efficiently. This is due to the availability of higher processing capabilities, via GPUs, and the large amount of data at disposal, i.e., Big Data from the internet~\cite{HintonNature}. The typical machine learning algorithms are divided into three classes. A first class of machine learning algorithms is named {\it supervised learning}~\cite{RussellAI}. These algorithms are trained via feeding them with labeled data, e.g., training a neural network with pictures of cats. After several iterations of the learning process, the network is able, if properly trained, to differentiate a picture from a cat from one of, say, a dog. A second class of machine learning algorithms is named {\it unsupervised learning}~\cite{RussellAI}. In this case, no labeled data is provided, while, in turn, the learning is produced via establishing correlations in the available data itself that one wants to classify. Via grouping this data in clusters, one is able to allocate new data inside one of these clusters, according to the highest proximity to it following some figure of merit, e.g., mathematical distance. Finally, a third class of machine learning algorithms, that is possibly more similar to the way the human brain itself learns, is the one of reinforcement learning~\cite{RussellAI,SuttonBarto}. In this kind of learning, a system, the {\it agent}, which can be, for instance, a robot, a computer program, a chemical molecule, or a quantum device, interacts with another system, the {\it environment}, which is the outer world to it, that is also relevant to its dynamics~\cite{RussellAI,SuttonBarto}. In this interaction, the agent obtains information from the environment, and also carries out some action on it, possibly modifying it. It then decides on a strategy, or policy, on how to proceed, depending on the feedback obtained from the interaction and action, and, iterating this procedure several times, aims at achieving some final goal. Reinforcement learning is the kind of learning that has taken more time to develop inside machine learning, but it has proven one of the most successful ones in the past few years, with the impressive achievements of AlphaGo and AlphaGo Zero in beating the Grandmasters of Go~\cite{AlphaGo,AlphaGoZero}. Go is a highly strategic game, much more complex than chess, which was not expected to be won by an artificial intelligence anytime soon. However, via a combination of supervised and reinforcement learning, AlphaGo beat some of the leading human experts at Go~\cite{AlphaGo}. Later on, AlphaGo Zero beat AlphaGo by 100-0, employing only reinforcement learning, playing millions of times against itself and learning in each iteration a bit more, an impressive achievement without any external training~\cite{AlphaGoZero}.

A different kind of approach is the concept of autoencoder~\cite{HintonNature}. It consists of a neural network that allows for compressing information, in the following fashion: the network contains an input layer (encoder), with a number of neurons, say, $n$. Then, it has one or more inner layers, with a number of neurons $n'<n$. And finally, it is connected to an output layer (decoder) with again $n$ neurons. One then trains this kind of network with data, assuming that it can be compressed (namely, it has common features). The process consists on aiming at matching the output signal of the network with the input signal, i.e., being able to employ only the $n'$ neurons to encode the information, instead of the $n$ initial ones, which is a larger number. If the process is successful, one may then discard the decoder and keep only the input layer, the encoder, therefore compressing the information.

In this Section, we will focus firstly on the machine learning algorithm known as reinforcement learning, and its recent development in the quantum realm, i.e., {\it quantum reinforcement learning}. We will describe the pioneering work by different groups in this topic, as well as our own results of the past few years.

Later on, we will analyze the quantum version of autoencoders, namely, {\it quantum autoencoders}, and describe the efforts by several groups both in theory and experiments in developing this technology.

\subsection{Quantum reinforcement learning}

\begin{figure}[h!]
\begin{center}
\includegraphics[width=\textwidth]{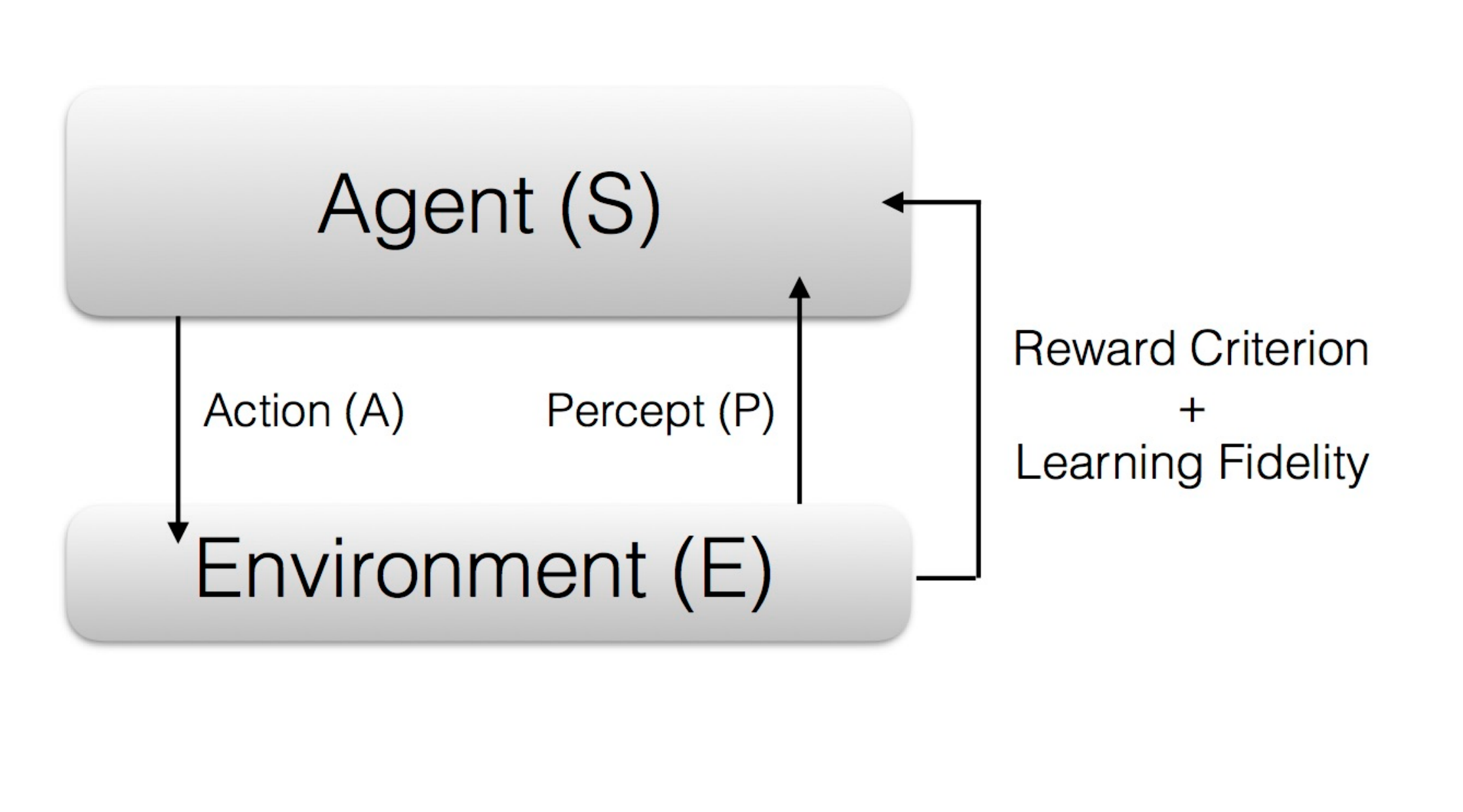}
\caption{Scheme of reinforcement learning. A system, called {\it agent}, interacts with the external world, the {\it environment}, realizing some action on it, as well as obtaining information from it, via a percept. Subsequently, the agent decides on a strategy on how to act or improve itself, by means of a reward criterion that depends on the action and percept, and the process is iterated. The motivation for the agent to follow this procedure is achieving a final goal. In the quantum reinforcement learning scenario, either agent, environment, or both, should be quantum, and they may exchange either quantum or classical information, or a mixture of both, with possible feedback in each iteration. Adapted from Ref.~\cite{LamataQRL}.}
\label{RLScheme}
\end{center}
\end{figure} 

In the field of quantum machine learning, quantum reinforcement learning (see Fig.~\ref{RLScheme}) is a paradigm that has emerged in recent years~\cite{DongQRL,PaparoPRX,DunjkoPRL,LamataQRL,CardenasQRL,AlbarranQRL,ShangYuQRL,AlbarranMLST,MarquardtRL,BukovPRX}. Some of the published works consider a quantum agent and a quantum environment (so called QQ scenario)~\cite{DunjkoPRL,LamataQRL,CardenasQRL,AlbarranQRL,ShangYuQRL,AlbarranMLST}, others consider a quantum agent and classical environment (QC)~\cite{DongQRL,PaparoPRX,DunjkoPRL}, another possibility would be to analyze a classical agent and quantum environment (CQ), while the fourth available combination (CC) corresponds to the purely classical situation of standard reinforcement learning.

In Ref.~\cite{DongQRL}, a mapping of classical reinforcement learning onto quantum reinforcement learning, connecting states of the former with quantum states of the latter, is proposed. Quantum superposition and entanglement are involved in the quantum algorithm, giving evidence of a possible resource gain with respect to classical reinforcement learning. The proposed protocol is based on Grover algorithm for unstructured search with quantum computers~\cite{Grover}. 

Ref.~\cite{PaparoPRX} analyzes a possible quantum advantage of a quantum agent interacting with a classical environment with classical channels, where the speedup would come from a faster quantum processor in the agent. This would be based, as in the previous case, in Grover search.

Ref.~\cite{DunjkoPRL} explores quantum reinforcement learning among other types of quantum learning, showing that quadratic improvements in learning efficiency, as well as exponential improvements in performance over limited amounts of time, may be achieved. For this, they consider a quantum agent and a quantum oracular environment.

The series of references~\cite{LamataQRL,CardenasQRL,AlbarranQRL,ShangYuQRL,AlbarranMLST} deal with quantum reinforcement learning in the QQ scenario, namely, quantum agent and quantum environment, in an increasing amount of complexity, and focusing on possible implementations. Refs.~\cite{LamataQRL,CardenasQRL} propose basic protocols of quantum reinforcement learning with superconducting circuits, including either coherent feedback inside the protocol, or projective measurements. In both cases, a single copy of the quantum environment state is considered, and the goal of the agent is to achieve a final quantum state that is identical to the environment one while, necessarily, modifying the latter. Ref.~\cite{AlbarranQRL} goes a step forward, taking into account several copies of the quantum environment state, such that the agent can achieve an increasing overlap with this state, which is its final goal. In Ref.~\cite{ShangYuQRL}, an experimental implementation of the previous proposal in quantum photonics was carried out. Interestingly, a gain of this quantum reinforcement learning protocol  with respect to standard quantum tomography was achieved, in the limited resources scenario. In the subsequent Ref.~\cite{AlbarranMLST}, an extension of the protocol to unknown operations, instead of states, was analyzed. Namely, this proposal would allow one to determine an unknown quantum operation, or, equivalently, its eigenvectors and eigenvalues, via a quantum reinforcement learning protocol.


\subsection{Quantum autoencoders}

\begin{figure}[h!]
\begin{center}
\includegraphics[width=\textwidth]{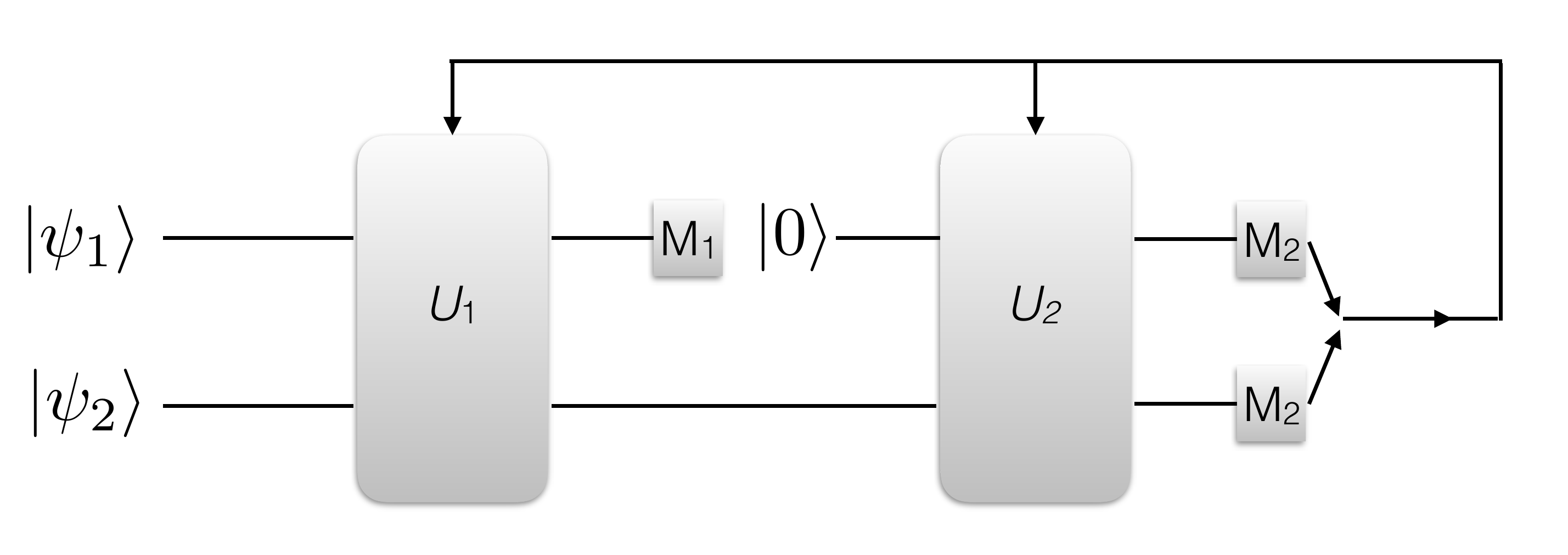}
\caption{Scheme of a quantum autoencoder, based on the algorithms in Refs.~\cite{AlanAutoencoder,KimAutoencoder}.  A set of input quantum states in a certain Hilbert space is encoded into a smaller Hilbert space via a learning process. Adapted from Ref.~\cite{LamataAutoencoder}.}
\label{Qauto1}
\end{center}
\end{figure}

Recently, proposals and experiments for quantum autoencoders have been explored by different groups (see Fig.~\ref{Qauto1}). In Refs.~\cite{AlanAutoencoder,KimAutoencoder}, one is given a set of states, inside a certain Hilbert space, and the aim is to be able to consider a smaller section of the Hilbert space that contains all the relevant information, namely, to somehow compress the quantum information onto a smaller amount of quantum bits or qubits. To this aim, in analogy with classical autoencoders, a quantum circuit composed of an input series of qubits is connected, via a parameterized unitary operation, to a smaller amount of qubits, and, subsequently, by means of a second parameterized unitary gate, to the same amount of qubits as the input. By feeding this quantum circuit with the set of states considered initially, and by measuring the output of the circuit, one may train the device, via classical feedback onto the parameters of the two unitary gates, in order to maximize the output fidelity with respect to the input states. This is similar to the standard autoencoder but in a quantum scenario. If the training is successful, then one may discard the final unitary gate, equivalent to the decoder, and keep only the initial one, the encoder, therefore reducing the amount of quantum information needed for practical applications. This could be useful, e.g., for quantum simulations~\cite{AlanAutoencoder}. An experimental realization of the previous proposal~\cite{AlanAutoencoder} has been carried out in a quantum photonics platform employing a three-level quantum system, a qutrit~\cite{TischlerAutoencoder}.

\begin{figure}[h!]
\begin{center}
\includegraphics[width=0.8\textwidth]{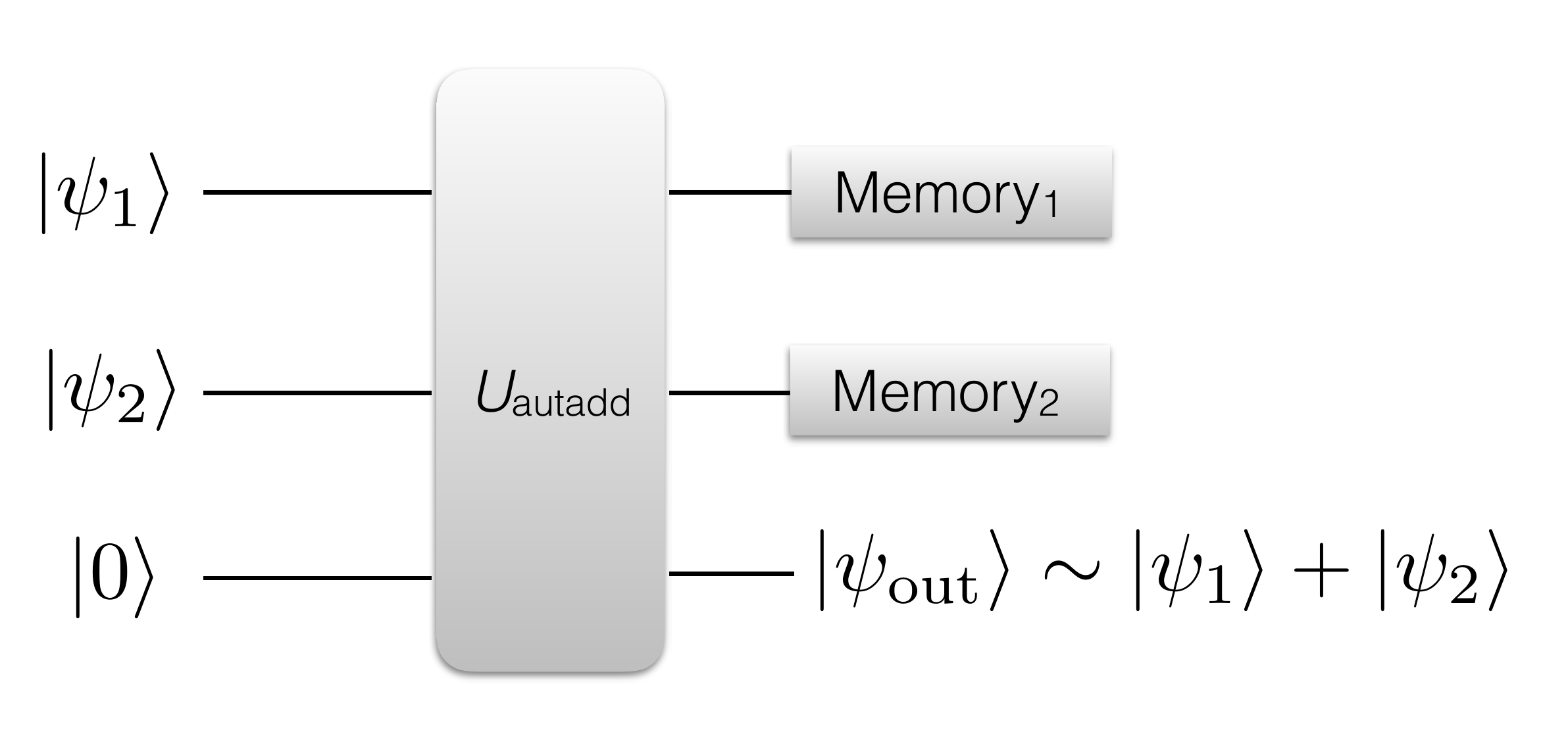}
\caption{Scheme of a quantum autoencoder based on approximate quantum adders. The two unknown input qubit states are encoded onto a single qubit, to a certain fidelity, via an approximate quantum adder operation. Adapted from Ref.~\cite{LamataAutoencoder}.}
\label{Qauto2}
\end{center}
\end{figure}

A different approach to quantum autoencoders was put forward in Refs.~\cite{LamataAutoencoder,DingAutoencoder} (see Fig.~\ref{Qauto2}). This is based on encoding the initial quantum information in a more reduced amount of qubits via the use of approximate quantum adders. In Ref.~\cite{UnaiAdder}, a theorem was proven that showed that a unitary operation able to add two unknown quantum states is forbidden by the laws of quantum physics (see also Ref.~\cite{Oszmaniec}). Nevertheless, this paper defined approximate quantum adders optimized according to diverse pre-specified criteria. In Ref.~\cite{RuiGenetic}, an optimization of approximate quantum adders based on genetic algorithms, a kind of machine learning algorithm, was realized. Subsequently, Ref.~\cite{LamataAutoencoder} developed a proposal for a quantum autoencoder based on the quantum adders optimized in Ref.~\cite{RuiGenetic}. Finally, Ref.~\cite{DingAutoencoder} implemented the previous proposal in the Rigetti cloud quantum computer, showing its feasibility.

Finally, another kind of autoencoders in the quantum realm has been proposed and implemented in the same article~\cite{AminAutoencoder}, namely, variational quantum autoencoders, which describe generative models implemented via quantum Boltzmann machines~\cite{AminMelkoQBM}.

\section{Quantum biomimetics\label{QAL}}

Quantum biomimetics is the field that aims at connecting other two previously disconnected research areas such as quantum technologies and biological systems~\cite{al1,al2}. This has links to quantum machine learning in areas such as, e.g., quantum neural networks, which are also biomimetic systems. Quantum artificial life is a field inside quantum biomimetics in which the objective is to design quantum individuals that can self-replicate in a quantum way, namely, in a compatible manner with the no-cloning theorem~\cite{qubiom1,qubiom2,qubiom3,WootersZurek}. A second field in quantum biomimetics is the one called quantum memristors~\cite{qmem1,qmem2,qmem3,qmem4}. These are inspired in the classical memristor, the fourth element in electronic engineering in addition to the resistor, capacitor, and inductor~\cite{DiVentraMemristors}. A classical memristor can process information at the same time as having a memory, and may provide a novel paradigm for computing, with a neuromorphic architecture. Equivalently, quantum memristors constitute a basic building block for a neuromorphic quantum processor. 

In this Section, we will first revise the topic of {\it quantum artificial life}, beginning with theoretical works~\cite{qubiom1,qubiom2} and finishing with an experimental implementation in the IBM cloud quantum computer~\cite{qubiom3}. 

Later on, we will describe the topic of {\it quantum memristors}, according to a series of results in the literature~\cite{qmem1,qmem2,qmem3,qmem4}.

\subsection{Quantum artificial life}

Artificial life is a research field that aims to reproduce biological properties common to living systems in artificially engineered systems~\cite{al1,alReview}. Some examples are self-assembling robots~\cite{aiSAR}, self-replicating chemicals~\cite{aiChemEvol} or computing programs~\cite{alPrograms}, as well as living neurons grown in the lab and connected to external devices~\cite{aiSyntBio}. One of the main areas inside this thriving field is the self-replicating systems. Living systems acquire complexity and structure via Darwinian evolution, or natural selection, and this is a combination of self-replication, introduction of variability via mutations as well as genome exchange, and selection of the fittest. Therefore, self-replication is one of the hallmarks of life. In the research line of {\it quantum artificial life}, a series of works has explored the possibility for elementary quantum systems to undergo self-replication~\cite{qubiom1,qubiom2,qubiom3} . This is a non-trivial issue given that the no-cloning theorem~\cite{WootersZurek} prevents the perfect copying of unknown quantum states. Thus, perfect self-replication onto distinct progeny individuals would not be allowed. What these works show is that the full quantum information can be faithfully transferred to the progeny individuals, under ideal conditions, if one maps the classical information of the diagonal elements of the parent density matrix onto the different progeny systems. On the other hand, the coherences of the parent density matrix would be mapped onto quantum correlations of the progeny systems. Therefore, entanglement plays a crucial role in the context of having self-replicating quantum systems.

\begin{figure}[h!]
\begin{center}
\includegraphics[width=\textwidth]{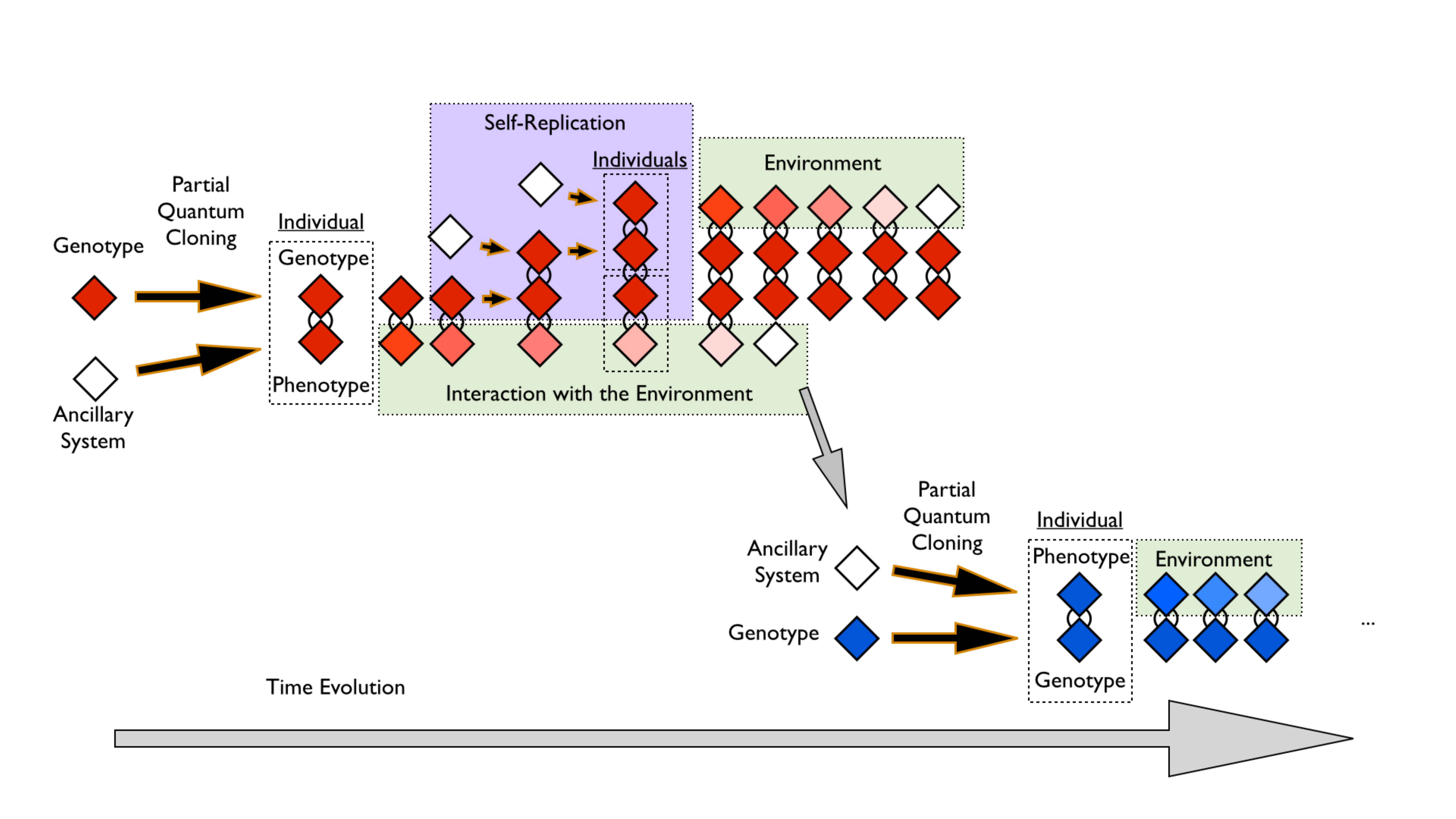}
\caption{Scheme for quantum artificial life. A quantum living unit, or quantum individual, is created from two qubits, which constitute the genotype and phenotype. The phenotype then decoheres when coupled with a quantum reservoir, until producing the ``death'' of the quantum individual. While still ``alive'',  the quantum individual can undergo self-replication, via coupling with further ancillary qubits. In order to transfer its full quantum information, the original quantum individual becomes entangled with its progeny quantum individuals. Adapted from Ref.~\cite{qubiom2}.}
\label{QbiomTheory}
\end{center}
\end{figure}

In Ref.~\cite{qubiom1}, a first analysis of biomimetic cloning of quantum observables was carried out, suggesting that the full quantum information may be transferred onto successive generations via cloning of classical information together with controlled transfer of quantum correlations. Later on, Ref.~\cite{qubiom2} proposed an implementation of a model of quantum artificial life. In this model, a quantum living unit, or quantum individual, is composed of two qubits, one for the genetic information, or genotype, and another for the expression of the genotype in the environment, or phenotype. The individual is created via a partial cloning operation of the genotype onto the phenotype via a Controlled-NOT operation, a kind of entangling two-qubit quantum gate. Subsequently, the phenotype can get coupled with a quantum reservoir, that induces decoherence on it. After a certain amount of time, the phenotype has almost totally decohered, and one considers that it has ``died'', at least as an artificial life individual, and according to this model. While the quantum individual is still ``alive'', it can couple with further ancillary qubits, carrying out additional partial cloning operations with Controlled-NOT gates, and produce new quantum individuals, which become entangled to the initial one, as previously explained (see Fig.~\ref{QbiomTheory}).

\begin{figure}[h!]
\begin{center}
\includegraphics[width=\textwidth]{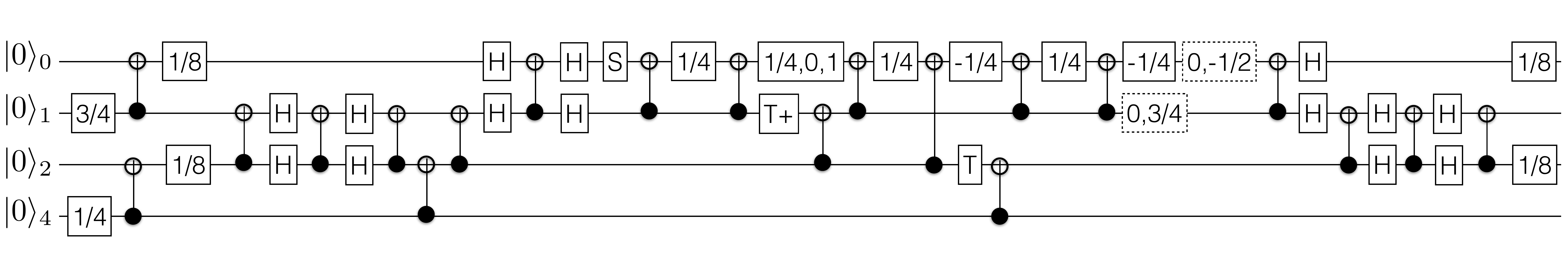}
\caption{Quantum circuit for the most complex quantum artificial life protocol experimentally carried out in Ref.~\cite{qubiom3}. The number of two-qubit Controlled-NOT entangling gates is of 20. H denotes a Hadamard gate, and the numbers inside the single-qubit gates indicate different phases, as described in Ref.~\cite{qubiom3}.}
\label{qcd3}
\end{center}
\end{figure}

\begin{figure}[h!]
\begin{center}
\includegraphics[width=\textwidth]{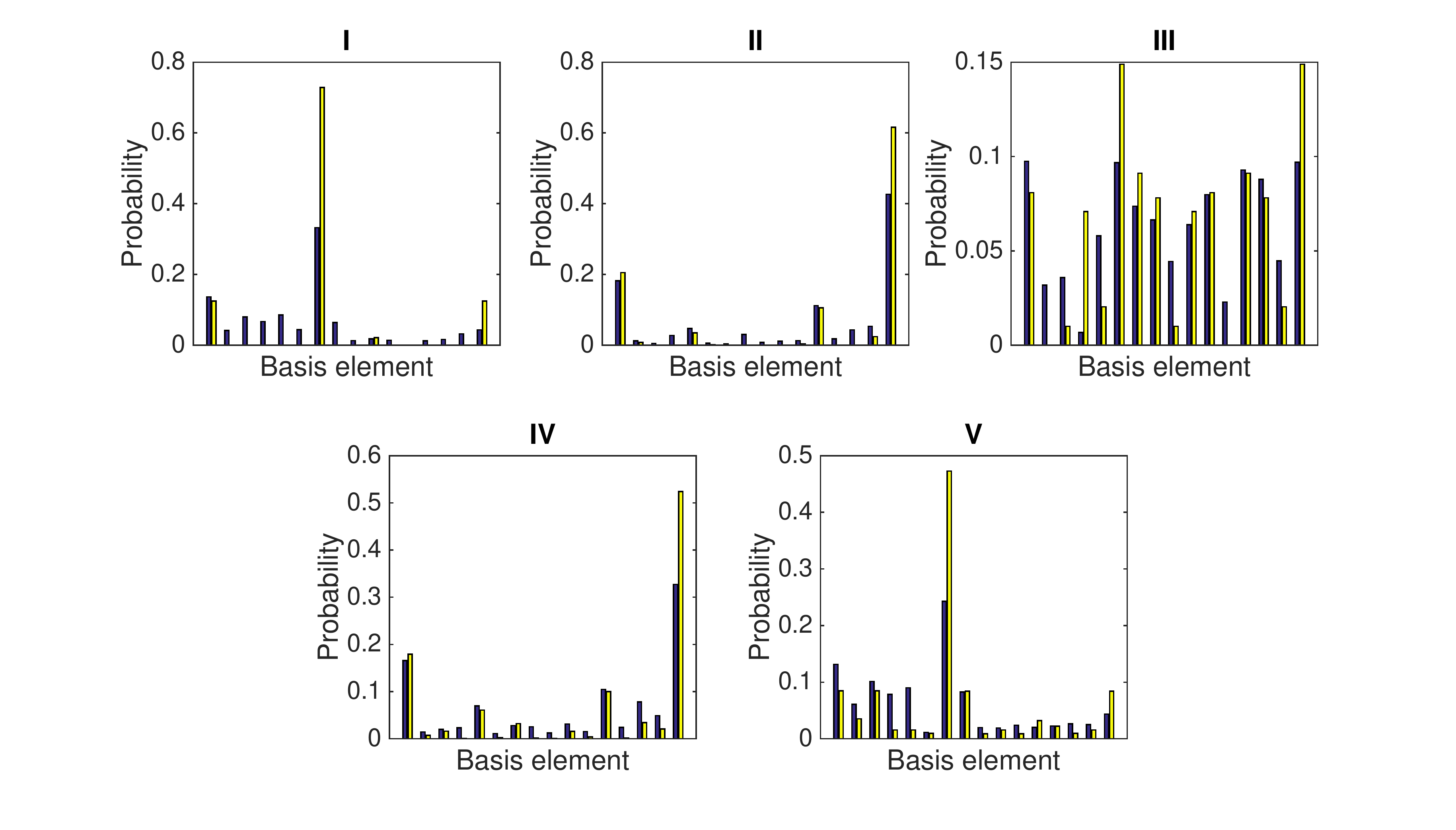}
\caption{Theoretical (yellow) and experimental (blue) data of the different experiments of a quantum artificial life system in the IBM cloud quantum computer. Adapted from Ref.~\cite{qubiom3}.}
\label{QbiomExp}
\end{center}
\end{figure}

In Ref.~\cite{qubiom3}, a quantum implementation of the quantum artificial life model of Ref.~\cite{qubiom2} was carried out in the IBM cloud quantum computer. Fig.~\ref{qcd3} shows the most complex quantum circuit that was implemented in this experiment. The number of entangling gates was of 20. On the other hand, Fig.~\ref{QbiomExp} depicts the theoretical (yellow) together with the experimental (blue) data of the different experiments of quantum artificial life carried out in the IBM cloud quantum computer. Even though the fidelities are not perfect, the experimental setup qualitatively reproduces the theoretical prediction to a large extent, for an experiment of a certain complexity.

\subsection{Quantum memristors}

The other topic in quantum biomimetics we will describe here is the one of quantum memristors~\cite{qmem1,qmem2,qmem3,qmem4}. In Ref.~\cite{qmem1}, a quantum version of a memristor was introduced, in which a two level quantum system encodes the quantum information, being supplemented with a weak measurement to acquire information, as well as feedback onto the system depending on the measurement outcome. It was proven that this device possesses hysteresis, and therefore a memory, in addition to processing capabilities. It was conjectured to be able to simulate non-Markovian quantum systems.

Subsequently, two proposals for implementations of quantum memristors in different quantum platforms were put forward, namely, superconducting circuits~\cite{qmem2} and quantum photonics~\cite{qmem3}. In the former, the memristive behavior emerges from quasiparticle-induced tunneling when supercurrent cancellation takes place. For realistic parameters, the appearing hysteretic behavior may be measured using state-of-the-art tomography of the phase-driven tunneling current. In the latter, the memristive behaviour arises from a tunable beam splitter, to which the input state is directed, together with feedback from one of the output ports onto the beam splitter that modifies its reflectivity depending on the output signal.

Finally, another result has been published on some complementary devices to quantum memristors, namely, qubit-based memcapacitors and meminductors~\cite{qmem4}.

As we have introduced, classical memristors constitute a paradigm for neuromorphic computing. Equivalently, quantum memristors represent a building block for a neuromorphic quantum device. A proven speedup with respect to their classical counterparts is still an open problem. Nevertheless, they represent a novel and promising quantum system that may play an important role in the future in a wide variety of scenarios as distributed quantum computing, quantum neural networks, and simulation of non-Markovian quantum systems.

\section{Conclusions\label{Conclusions}}

In this Perspective, we have given an overview of topics inside the fields of quantum machine learning and quantum biomimetics. In quantum machine learning, we have explored the topics of quantum reinforcement learning, on the one hand, and quantum autoencoders, on the other hand. In quantum biomimetics, we have described the results in quantum artificial life and quantum memristors.

Quantum reinforcement learning considers an agent interacting with an environment, obtaining information from it, as well as acting on it. The agent, typically a quantum system, can follow a policy to achieve a predetermined goal. Different works in the literature show possible resource gains of quantum reinforcement learning with respect to its classical counterpart, as well as describe feasible experimental realizations in controllable quantum platforms.

Quantum autoencoders are a kind of quantum machine learning protocol in which the aim is to reduce the amount of quantum resources needed to encode a certain set of quantum states. Via a training process, with a certain parameterized quantum circuit, one may compress the quantum information from a certain Hilbert space onto a smaller space of states. A further variant of quantum autoencoders proposed in the literature is the one based on approximate quantum adders. These would allow for encoding a certain number of quantum states on a reduced number of quantum states via previously-optimized approximate quantum adders. Both kinds of quantum autoencoders have been realized experimentally. The first one in quantum photonics, while the second one in superconducting circuits.

Quantum artificial life is a novel paradigm inside quantum biomimetics that may enable a quantum system to self-replicate and propagate its quantum information to further quantum systems. An open problem that remains to be addressed in these systems is whether one may be able to encode optimization problems by allowing the quantum individuals to compete, in some kind of quantum Darwinian evolution.

Quantum memristors are quantum devices with processing capabilities and memory, and constitute the basic building blocks for neuromorphic quantum computing. They may be employed in the future for distributed quantum computing, quantum neural networks, as well as quantum simulation of non-Markovian quantum systems.

Summarizing, quantum machine learning and quantum biomimetics are two emerging and promising fields inside quantum technologies, which may enable a wide variety of applications in the future.The connection between both fields is apparent in the sense that living systems, either natural or artificial, have often the ability to learn. With the advent of Noisy Intermediate-Scale Quantum (NISQ) computers~\cite{NISQ}, an enormous amount of computing power will be available, with desirable properties of speed and low energy consumption. Only the future can tell which progress may these technologies bring about.

\section*{Acknowledgements} I acknowledge collaborations and useful discussions on the topics presented in this article with Enrique Solano, Mikel Sanz, Unai Alvarez-Rodriguez, Jos\'e Mart\'in-Guerrero, Pablo Escandell-Montero, Francisco Albarr\'an-Arriagada, Juan Carlos Retamal, Francisco C\'ardenas-L\'opez, Yongcheng Ding, Xi Chen, Shang Yu, and Rei Li. Funding from Spanish PGC2018-095113-B-I00 (MCIU/AEI/FEDER, UE) is acknowledged.

\section*{Data Availability Statement}

Data sharing does not apply to this article as no new data were created or analysed.

\end{document}